\documentclass[nofootinbib,aps,prl,longbibliography,twocolumn]{revtex4-2}
\usepackage{placeins}

\usepackage{amsmath,amssymb,amsfonts,amsthm,bm}
\usepackage{color,xcolor,dsfont}
\usepackage{graphicx,esint,relsize,comment,float}
\usepackage[titletoc,toc,title]{appendix}
\usepackage[english]{babel}
\usepackage{tgtermes}  

\usepackage{hyperref}
\usepackage{cleveref}
\crefname{section}{Section}{Sections}
\usepackage{lipsum}
\usepackage{ulem}

\newcommand{\abs}[1]{\left\vert #1 \right\vert}

\begin{document}

\title{\bf{The Diffusive Nature of Housing Prices}}

\author{Antoine-Cyrus Becharat$^{1,2}$}
\email{antoinecyrus.becharat@gmail.com}
\author{Michael Benzaquen$^{1,2,3}$}
\author{Jean-Philippe Bouchaud$^{1,3,4}$}

\affiliation{$^1$Chair of Econophysics and Complex Systems, École Polytechnique, 91128 Palaiseau Cedex, France}
\affiliation{$^2$LadHyX UMR CNRS 7646, École Polytechnique, 91128 Palaiseau Cedex, France}
\affiliation{$^3$Capital Fund Management, 23 Rue de l’Université, 75007 Paris, France}
\affiliation{$^4$Académie des Sciences, 23 Quai de Conti, 75006 Paris, France}

\date{\today}

\begin{abstract}
We analyze the French housing market prices in the period 1970-2022, with high-resolution data from 2018 to 2022. The spatial correlation of the observed price field exhibits logarithmic decay characteristic of the two-dimensional random diffusion equation -- {\it local} interactions may create {\it long-range} correlations. We introduce a stylized model, used in the past to model spatial regularities in voting patterns, that accounts for both spatial and temporal correlations with reasonable values of parameters, some fitted on impulse response data. Our analysis reveals that price shocks are persistent in time and their amplitude is strongly heterogeneous in space. Our study quantifies the diffusive nature of housing prices that was anticipated long ago \cite{clapp1994, Pollakowski1997}, albeit on much restricted, local data sets. 
\end{abstract}

\maketitle


Complex spatial patterns often result from a subtle interplay between random forcing and diffusion, like for example surface growth \cite{barabasi1995fractal} or fluid turbulence \cite{frisch1995turbulence}. One can also expect such competition between heterogeneities and diffusion to take place in socio-economic contexts. For example, word of mouth leads to spreading of information or of opinions. Provided the spreading mechanism is local enough (i.e.~before the advent of social media), the large scale description of such phenomena is provided by the diffusion equation that leads to specific predictions for the long-range nature of spatial correlations of voting patterns, which seems to be validated by the analysis of empirical data \cite{Borghesi2010, borghesi2012election, fernandez2014voter}.

\begin{figure}
    \centering
    
    \includegraphics[height=0.46\linewidth]{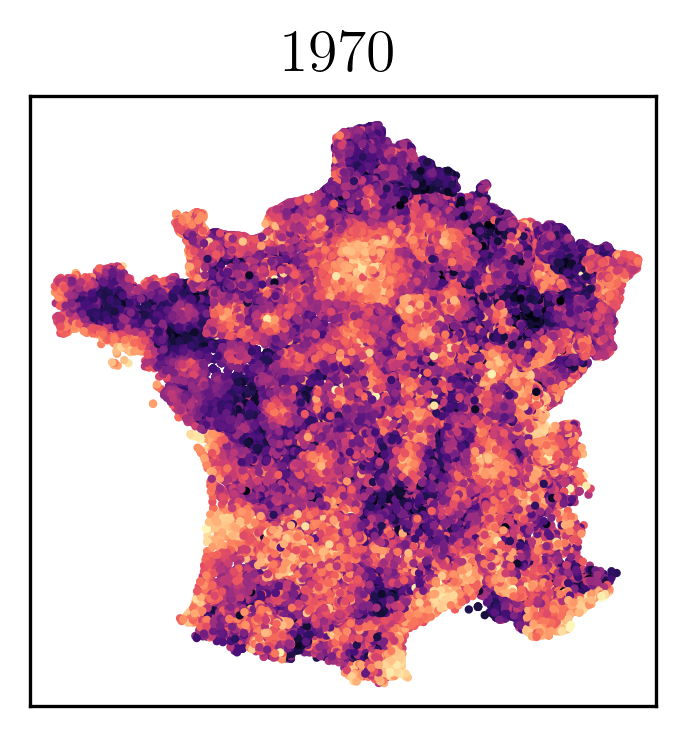} 
    \includegraphics[height=0.46\linewidth]{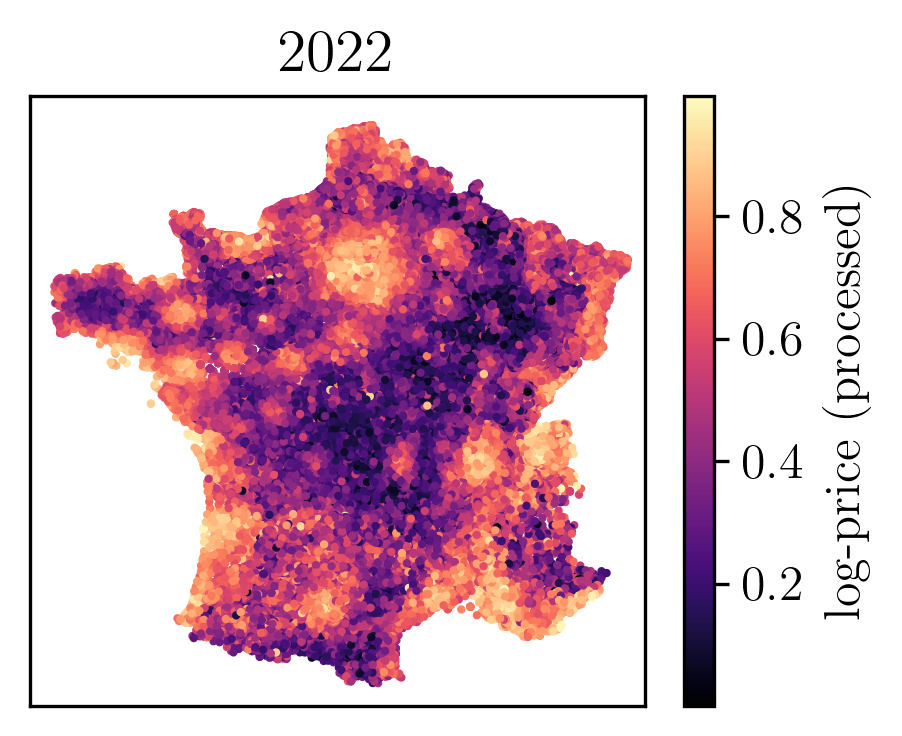} 
    \caption{Spatial transaction log-prices $p$ distribution in France in 1970 (left) and in 2022 (right). We use a sigmoid transformation of the log prices rescaled by their mean and divided by their standard deviation in order to highlight price differences. As seen in this plot, high prices are concentrated around France's principal cities and on the coasts and mountains, but the price pattern clearly displays spatial diffusion. Data from \cite{Piketty2023}.}
    \label{fig:map France}
\end{figure}

One may argue that housing prices should display similar patterns. Indeed, it is intuitively clear that the price of real estate in a given district is affected, among many other factors, by the price of the surrounding districts, through a sheer proximity effect. This is enough to generate a diffusion term in any coarse-grained description of the spatio-temporal evolution of prices -- see below and SM-1 of the Supplemental Material \ref{sec:SM1} for more precise statements. The aim of this work is to present such a phenomenological description of the dynamics of the {\it price field} in a given region of space, and to compare analytical prediction to empirical data using spatially resolved transaction prices in France for the period 1970 to 2022 -- see Fig. \ref{fig:map France} for a visual representation of the price field that motivates our analysis. We will find what we consider to be rather remarkable agreement with theory, in view of the minimal amount of modeling ingredients. In particular, the logarithmic dependence of spatial correlations, characteristic of two-dimensional diffusion, is clearly visible in the data at all scales (see Fig. \ref{fig:msd spatial} below).

Due to its potent macroeconomic and systemic risk implications, housing prices have long been studied by economists, see \cite{geanakoplos2012getting}. One of the most famous description of the housing market is through the Hedonic prices hypothesis (see  e.g. \cite{Rosen1974}), which states that {\it goods are valued for their utility-bearing attributes. Hedonic prices are defined as the implicit prices of attributes and are revealed from observed prices of differentiated products and the specific amounts of characteristics associated with them}. In essence, we shall argue that real-estate prices in the vicinity of a given location is one of these characteristics. 

There is also a great body of empirical literature highlighting the links between the housing market prices and, for example, violence \cite{Besley2012} or school grades \cite{Figlio2004}. This has naturally led to models of the housing market using reasonable assumptions. In particular, recent agent-based models of the housing market have been designed to explain price dynamics \cite{geanakoplos2012getting}, or its link with social segregation. Ref.~\cite{Feitosa2008} observed that segregation patterns can be observed even with the simplest parameter setting in an agent-based model of the housing market. Ref.~\cite{Pangallo2019} showed how such models could be very helpful to test and apply effective policies to prevent social/racial segregation, in the same vein as Ref. \cite{Baptista2016} where the effectiveness of macro-prudential policies is tested on an agent-based model of the UK housing market. Interestingly, \citep{gauvin_modeling_2013} showed that social segregation is also strongly linked with social influence. 

Concerning spatial patterns, {\it local} studies from the mid-1990's have suggested the potential importance of spatial diffusion effects. For example, Clapp \& Tirtiroglu \cite{clapp1994} find evidence of local price diffusion from their empirical study of the metropolitan of Hartford, Connecticut. Pollakowski \& Ray \cite{Pollakowski1997} confirms these results at the local level, and conclude that housing prices are inefficient: {\it If housing markets were efficient, [...] shocks would either be confined to one area, in which case information transfer is irrelevant, or affect a number of areas, in which case the price changes should occur nearly simultaneously, not one after another.} These authors also note that price changes are auto-correlated in time (a feature that we will explicitly include in our theoretical model), which is a further sign of price inefficiency. Indeed, properly anticipated prices should not be predictable \cite{samuelson2016proof}. 

As we argue below, such {\it local} diffusion of prices is expected to create {\it long-range correlations} in the price field both in space and in time, which we observe in the data, {Although the presence of spatial correlations was noticed in \cite{Thibodeau1998}, their analysis did not provide any theoretical framework to elucidate the origin of these correlations. In particular, no mention was made of their long-range character, let alone of their specific logarithmic dependence that we establish empirically and justify theoretically in the following.} Other socio-economic variables are known to be long-range correlated \cite{borghesi2012election}, with far-reaching consequences on the statistical significance of many results in spatial economics, as forcefully argued in  \cite{kelly2019}.

Our theoretical framework aims at modeling the dynamics of the housing price field in a similar spirit as for the dynamics of opinions or intentions  \cite{schweitzer2000modelling, schweitzer2004coordination, Borghesi2010, bouchaud2014emergence}, or of criminal activities \cite{short2010}. We introduce a two-dimensional field $\psi({\bf r},t)$ which represents the deviation from the (possibly time dependent) mean of the log-price of housing around point ${\bf r}$ at time $t$. We then posit that such a field evolves in time according to the following  stochastic partial differential equation 
\begin{equation} \label{eq: diffusion equation}
    \frac{\partial \psi({\bf r},t)}{\partial t} = D \Delta \psi({\bf r},t) - \varkappa \psi({\bf r},t) + \eta ({\bf r},t) + \xi({\bf r}),
\end{equation}
where $\Delta$ is the Laplacian operator, $D$  a diffusion coefficient, $\varkappa$ a mean-reversion coefficient, $\eta ({\bf r},t)$  a Langevin noise with zero mean and short range time and space correlations, and $\xi({\bf r})$  a static random field with zero mean and short range correlations. The correlators of these terms are assumed to be of the following type:
\begin{align} \nonumber
    \left \langle \eta({\bf r},t) \eta({\bf r}',t') \right \rangle &= \frac{A^2}{T a^2} e^{-|t-t'|/T} g_a(|{\bf r} - {\bf r}'|); \\
    \left \langle \xi({\bf r}) \xi({\bf r}') \right \rangle &= \frac{\Sigma^2}{a^2} g_a(|{\bf r} - {\bf r}'|),
\end{align}
where $g_a(r)$ is a  bell-shaped function that decays over length scale $a$, such that $2 \pi \int_{r > 0} g_a(r) r {\rm d}r = a^2$. Note that in terms of dimensions, $[A^2]=[D]=[L^2 T^{-1}]$, $[\varkappa]=[T^{-1}]$ and $[\Sigma]=[L T^{-1}]$. 

The four different terms of Eq.~\eqref{eq: diffusion equation} capture the following features: (i) the diffusion term describes the proximity effect alluded to in the introduction and documented in Refs. \cite{clapp1994, Pollakowski1997}: pricey districts tend to progressively gentrify;
conversely, rundown districts lower the market value of their surroundings. (A more technical version of this argument is given in SM-1 \eqref{sec:SM1} . (ii) The mean-reversion term can be seen as a coupling between local log-prices and the mean log-price, here set to zero, and can be thought of as the result of long-range economic forces that keep prices within a country more or less in sync through the effect of e.g. migrations, policies or wealth inequalities. (iii) The time-dependent noise term $\eta$ models all idiosyncratic shocks affecting the ``hedonic'' variables determining the price of properties -- for example the creation of a local metro or train station, of a pedestrian zone, or adverse shocks like increase in local crime, floods, etc. The impact of such shocks is often drawn out in time, so we assume $\eta$ to be auto-correlated with a decay time $T$, in line with the observations reported in  \cite{Pollakowski1997}. (iv) The time-independent stochastic term $\xi$ is meant to represent persistent biases in the local quality of life in different regions, due to e.g. geographical features (close to the sea-shore, or to river banks, etc.). For simplicity, We have assumed that the spatial correlation lengths of both $\eta$ and $\xi$ are equal to the same value $a$. 

Now, Eq.~\eqref{eq: diffusion equation} makes detailed predictions for the spatial and temporal correlations of the field $\psi({\bf r},t)$. To wit, the spatial variogram $\mathbb{V}(\ell,0) := \langle (\psi({\bf r},t) - \psi({\bf r}',t))^2 \rangle_{|{\bf r} - {\bf r}'| = \ell}$ can be explicitly computed in the range $\max(a, \sqrt{DT}) \ll \ell \ll \ell^\star$ (where $\ell^\star:=\sqrt{D/\varkappa}$), and reads (see SM-2.2 \cite{supplemental_material}):
\begin{equation} \label{eq:variogram}
    \mathbb{V}(\ell,0) \approx  \frac{A^2}{2 \pi D}\log \frac{\ell}{\ell^\star} - \frac{\Sigma^2}{4 \pi D^2} \ell^2 \log \frac{\ell}{\ell^\star} + C,
\end{equation}
where $C$ is a constant. Note that the first term is the familiar logarithmic correlation of the Gaussian free-field in two dimensions, see e.g. \cite{edwards1982surface}. For $\ell \gtrsim \ell^\star$, the variogram reaches a plateau value. 

Similarly, the temporal variogram $\mathbb{V}(0,\tau) := \langle (\psi({\bf r},t) - \psi({\bf r},t + \tau))^2 \rangle$ can be computed, but the final expression is cumbersome and depends on the relative position of three time scales: $\varkappa^{-1}$, the correlation time $T$ and the typical diffusion time $S = a^2/D$ over length scale $a$, see SM-2.3 \eqref{sec:SM2} . There are typically four regimes, a short time regime where $\mathbb{V}(0,\tau) \propto \tau^2$ that reads
\begin{equation} 
    \mathbb{V}(0,\tau) = \frac{ A^2}{16 \pi D} \log\left(\frac{1+\frac{T}{S}}{1+\varkappa T}\right) \, \frac{\tau^2}{T^2}, \quad \tau \ll T,S
\label{eq:variogram temporel tau petit}
\end{equation}
followed by two intermediate regimes where $\mathbb{V}(0,\tau) \propto \tau$ and $\log \tau$, and finally a saturated regime for $\varkappa \tau \gg 1$.

In the next sections, we will compare these predictions to empirical data, with good overall agreement. We will find that the spatial variogram is well described by a pure logarithm, i.e. the first term of Eq. \eqref{eq:variogram} -- this allows us to determine the ratio $A^2/D$. With the same value of $A^2/D$, we then fit the temporal variogram with reasonable values of $T$ and $S$.

We conducted extensive empirical analyses based on two data sources. The first one is accessible online via the DVF (Demande de Valeur Foncière) website, and displays every housing market transaction in France between 2018 and 2022. This data include the price of the property, its surface and its spatial coordinates. This allows us to study both transaction prices and prices per square meter, up to the granularity of a given point in space.
The second data source comes from \cite{Piketty2023}, where the authors compiled a wealth of socio-economic indicators, spanning from 1970 to 2022 \footnote{
For the specific case of the housing market. Other socio-economic indicators cover an even longer time span. We in fact found similar logarithmic correlations for, e.g., the alphabetization rate in France.}, including housing market prices, but the dataset only contains average transaction prices per \textit{communes} in France up to 2022 and average prices per square meter per \textit{communes} from 2014 to 2022. \footnote{The housing market data compiled by \cite{Piketty2023} for the years 2014-2022 comes from the DVF database, and is averaged per \textit{communes}.} 
Even though the latter data source is much less granular than the DVF dataset, its  time span of 52 years allows us to investigate the temporal variogram of prices, and so this is the data we focus on in the main text. (The DVF data only span 5 years, which will turn out to be of the same order of magnitude as the correlation time $T$ of the noise). For empirical findings on prices per square meters from DVF, which are fully consistent with those established for transaction price, see SM-3 \eqref{sec:SM3} .  

We first show a color map of transaction log-prices $p:=\log P$ across France in Figure \eqref{fig:map France},  sourced from \cite{Piketty2023}, to compare the spatial distribution of prices in France over the past five decades, a key aspect of our investigation. Indeed, one can see that the price distribution in France is far from uniform, and reveals spatial diffusion around big cities, coastal regions or ski resorts.

\begin{figure}
    \centering
        \includegraphics[width=\linewidth]{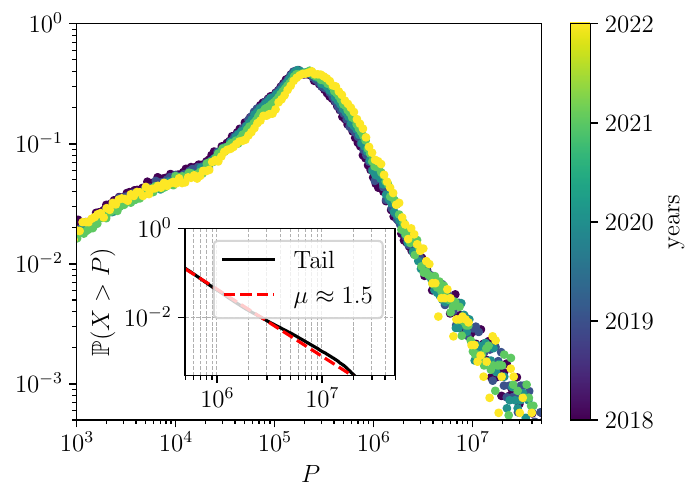} 
    \caption{Distribution of all transaction log-prices ${p}:=\log {P}$, for the 5 years of DVF data. Note the double hump shape, reflecting a mixture of two distributions, corresponding to prices in cities and prices in the countryside. The right tail, for property prices above $500,000$ Euros, corresponds to a power-law tail for prices as $P^{-1-\mu}$ with $\mu \approx 1.5$.}
    \label{fig:distribution log price}
\end{figure}

Then, it is interesting to study the distribution of {\it individual} transaction log-prices $p$, unconditionally over the whole of France. Using the DVF data base, we find that the distribution of prices has a double hump shape, probably reflecting the superposition of two different price distributions for cities and for the countryside, see Fig. \ref{fig:distribution log price}. We show in SM-3, Fig. 2 \eqref{sec:SM3} , a comparison between the distribution of prices in the \textit{département} of la Creuse (chosen to represent a typical countryside district) and in Paris, highlighting the mixture of two distributions seen in the global price distribution for the whole of France.  The tail of the distribution of the transaction prices decays as $P^{-1-\mu}$ with $\mu \approx 1.5$, implying that the variance of the transaction prices is mathematically infinite. This should be compared to the Pareto tail of the wealth distribution in France, which decays with a similar exponent \cite{bach2015top}. The distribution of prices per square meters does not have the same shape, but has again a similar power-law tail, as shown in SM-3, Fig. 3 \eqref{sec:SM3}.  

 We now shift our focus to the spatial correlations of the logarithm of prices, which we characterize by the equal-time variogram $\mathbb{V}(\ell,0)$ defined above. The square-root of this quantity measures how different the (log-)prices are when considering two properties a distance $\ell$ away.\footnote{The spatial structure of transaction prices per square meters is investigated in SM-3, Fig. 1 \eqref{sec:SM3} .} We studied this quantity inside cities, {\it départements}, {\it régions} and the whole of France, with a different coarse-graining scale for the elementary cells over which we average the transaction prices $P$ in order to define the log-price field $p({\bf r})$. We choose hexagonal cells of area $0.73$ km$^2$ for the 17 cities considered,\footnote{
This leads, for instance, to the division of Paris into 185 neighborhoods.
} $5$ km$^2$ for {\it départements}, $30$ km$^2$ for {\it régions}, and $250$ km$^2$ for France. The results are shown in Fig. \ref{fig:msd spatial}. At all scales, we observe a logarithmic dependence on $\ell$, provided $\ell$ is smaller than the size of sector considered (see further down). Furthermore, the slope predicted by Eq.~\eqref{eq:variogram} is the same at all scales and equal to $A^2/2\pi D \approx 0.19$. The measured (log-)slopes of the variograms are extremely stable over the period 2018-2022 spanned by the DVF data. The other data source \cite{Piketty2023} allows one to measure the spatial variogram over a much longer history. However, the data collection and averaging procedures used in \cite{Piketty2023} seem to induce distortions in the price variograms when compared to the raw DVF data, that we do not fully understand. Still, the analysis of these variograms reveals that the slope of the short-distance logarithmic behavior is only weakly time dependent, before {saturating} for $\ell \approx 70$ km in 1970 and $300$ km nowadays, as seen in  SM-3, Fig. 4 of \eqref{sec:SM3} . A possible interpretation is that this crossover length is set by $\ell^\star = \sqrt{D/\varkappa}$ which has increased with time, either because $D$ has increased (faster spatial propagation of price changes) or because $\varkappa$ has decreased, reflecting larger wealth inequalities that allows for larger price dispersion, or both.

In order to determine the order of magnitude of the diffusion constant $D$ we analyse the propagation of price ``shocks'' induced by the opening of a TGV ({\it Train à Grande Vitesse}) train station in various cities, as shown in Fig. \eqref{fig:empirical D}. We find that $D$ is of order $50$ km$^2$/year, corresponding to prices adapting to a local shock on a scale of $7$ km after a year. 
This leads to a value of $A^2 \approx 2 \pi \times 0.19 D \sim 60$ km$^2$/year. We will comment on this value below, after having discovered that the noise amplitude $A^2$ is in fact space dependent.  

\begin{figure}
    \centering
    \includegraphics[width=\linewidth]{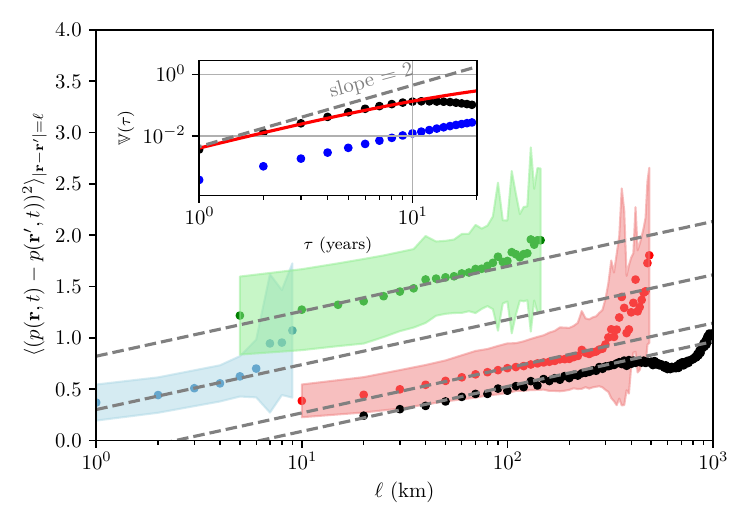}
    \caption{Spatial variogram for the log price field ${p}({\bf r})$ averaged over the period 2018-2022 for France as a whole, its \textit{régions}, \textit{départements} and cities, with their respective cross-sectional variability highlighted in shaded colors and the averages for each scale as filled circles. The black dashed lines have a slope equal to $0.19$ for all scales, corresponding to $A^2/D=1.2$. The different off-sets in the $y$ direction corresponds to the measurement noise contribution to the empirical field $p({\bf r})$. {The inset shows the comparison between $\mathbb{V}_1(\tau)$ and $\mathbb{V}_2(\tau)$ for the empirical data, in a log-log representation. We also show (in red) the fit found for $\mathbb{V}_1(\tau)$ with our theoretical equation. Note that the short time behavior of $\mathbb{V}_1(\tau)$ is in-between $\tau$ and $\tau^2$, indicating a non-zero correlation time $T$. We find $T=3.5$ years, $S=1$ year, $D=50 \text{km}^2$ per year and $A^2/D \approx 1.2$. The observed shift between $\mathbb{V}_1(\tau)$ and $\mathbb{V}_2(\tau)$ is a consequence of strong spatial heterogeneities, see SM-3, Fig. 5 \eqref{sec:SM3} . Note that with 50 years of data, only the first 10 years of lags are reliable. } }
    \label{fig:msd spatial}
\end{figure}

The reader must have noticed that although the {\it slopes} of the variograms are the same at all scales, they are shifted up and down in the y-direction. This is expected if one accounts for measurement noise. Indeed, the ``true'' price field $p({\bf r},t)$ is approximated here by an empirical average over the chosen cells of transaction prices. The larger the cell size and the smaller the dispersion of prices within each cell, the smaller such idiosyncratic contributions to the difference of prices for two neighboring cells.

Finally, note that the spatial variograms do not seem to reveal any departure from the $\log \ell$ behavior predicted by the first term of Eq.~\eqref{eq:variogram}, except at large distances where finite size and boundary effects start playing a role. Comparing the two terms of Eq.~\eqref{eq:variogram}, one concludes that the second term remains negligible provided $\ell \lesssim D/\Sigma$. Choosing $D = 50$ km$^2$/year and $\ell=500$ km, this holds provided $\Sigma \lesssim 0.1$, i.e. whenever idiosyncratic effects lead to persistent differential of price variations of at most $10 \%$ after a year and over $1$ km. We believe that this is indeed an upper bound to such idiosyncratic effects. 

Turning to the temporal variogram of prices, there are two different empirical definitions for such an object, which should lead to similar results if the system is (statistically) spatially homogeneous. One ($\mathbb{V}_1(\tau)$) is to compute the temporal variance of local price changes $p({\bf r},t) - p({\bf r},t+\tau)$ over the full time period, which is then averaged over ${\bf r}$. The second ($\mathbb{V}_2(\tau)$) is to remove from $p({\bf r},t)$ the spatial average of the log-price at time $t$, i.e. $\bar{p}(t)=\langle p({\bf r},t) \rangle_{{\bf r}}$, and then compute the average of $[p({\bf r},t) - \bar{p}(t) - (p({\bf r},t+\tau) - \bar{p}(t+\tau))]^2$ over both $t$ and ${\bf r}$. For a statistically homogeneous system, these two procedures lead to comparable results. However, as shown in Fig. \ref{fig:msd spatial}, our data reveals strong differences between $\mathbb{V}_1(\tau)$ and $\mathbb{V}_2(\tau)$, which can be accounted for by assuming that the variance $A^2$ of the driving noise $\eta$ is space dependent: $A^2 \to A^2({\bf r})$. In this case, spatial correlations lose their translation invariance but if one insists on computing them as a function of $\ell = |{\bf r} - {\bf r}'|$, one recovers Eq. \eqref{eq:variogram} with $A^2$ replaced by its spatial average
$\langle A^2 \rangle_{{\bf r}}$, see SM-3, Fig. 5 \ref{sec:SM3}.

\begin{figure}
    \centering
    \includegraphics[width=\linewidth]{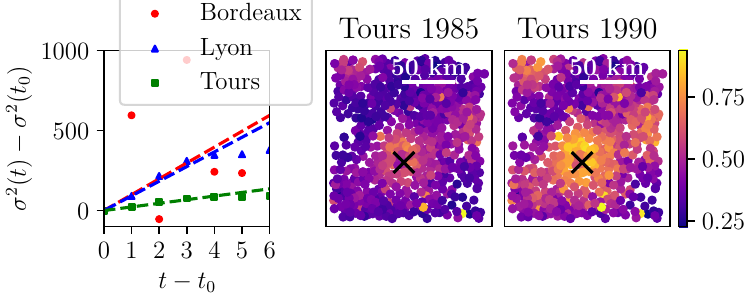}
    \caption{{Left: Impulse response function. We measure how housing prices diffuse around a newly opened TGV station establishing a fast train connection to Paris at time $t_0$ from Lyon, Tours and Bordeaux (see SM-4 \eqref{sec:SM4}  for a precise definition of $\sigma^2$). The x-axis is in years and the y-axis in km$^2$. From the slope of the linear initial regime we infer the diffusion constant $D$: $D \approx 90 \text{km}^2/\text{year}$ for Lyon, $D \approx 22 \text{km}^2/\text{year}$ for Tours and $D \approx 100 \text{km}^2/\text{year}$ for Bordeaux, roughly in agreement with our selection of $D\approx 50 \text{km}^2/\text{year}$ for the whole of France. Middle and right: visual representation of the price field centered around the city of Tours, marked by a black cross. The isotropic diffusion of the price field after the opening of the TGV train station in 1985 is clearly observable.}}
    \label{fig:empirical D}
\end{figure}

Now, it turns out that in the presence of spatial heterogeneities, the temporal variogram $\mathbb{V}_1(\tau)$ is also given by Eq.~\eqref{eq:variogram temporel tau petit} with $A^2 \to \langle A^2 \rangle_{{\bf r}}$, see SM-3, Fig. 5 \eqref{sec:SM3} . Hence we focus our attention to $\mathbb{V}_1(\tau)$ and attempt to fit it with our theoretical formula (see SM-2.3 \eqref{sec:SM2} ) with $T,S$ as adjustable parameters, with $\langle A^2 \rangle_{{\bf r}}/D$ fixed and set to $1.2$, close to the value inferred from spatial variograms. ($D$ itself has negligible influence on the goodness-of-fit, only the ratio $A^2/D$ matters). The optimal values are then found to be $S=1$ year, corresponding to a correlation length for shocks {$a = \sqrt{DS}= 7$} km, and a correlation time of $T=3.5$ years, such that $\sqrt{DT} = 13$ km. The order of magnitude of $A^2$ is expected to be $a^2/T \sim$ 30 km$^2$/year, a factor two times smaller than expected if $D = 50$ km$^2$/year, but not unreasonable in view of the crudeness of our model and the possibility to change the value of parameters without substantially affecting the joint goodness-of-fit of spatial and temporal variograms. For example, choosing $\langle A^2 \rangle_{{\bf r}}/D  =1.3$ leads to $T=S=2.5$ years and in this case $a^2/T \sim$ 50 km$^2$/year. 
Note that the short-time regime of $\mathbb{V}_1(\tau)$ is a sign that price changes are persistent, which is inconsistent with the hypothesis that the housing market is ``efficient'' \cite{Pollakowski1997}. In view of the large transaction costs incurred when buying a house, this is hardly surprising.

Finally, in order to account for the empirical difference between the two temporal variograms $\mathbb{V}_1(\tau)$ and $\mathbb{V}_2(\tau)$, one needs to introduce rather strong spatial heterogeneities in the noise amplitude $A^2$, that must vary by a factor of $10$ depending on the considered region, see SM-3, Fig. 5 \eqref{sec:SM3} . This is not very surprising in view of the very different structure of the housing market in international cities like Paris or Nice and the remote, low density regions like Lozère. A generalized version of our model, Eq.~\eqref{eq: diffusion equation}, that properly accounts for geographical heterogeneities that make both $D$ and $A^2$ space dependent, would however require a different, much more granular calibration strategy.

In conclusion, we have proposed a simple, general dynamical model for the spatial evolution of housing prices inspired from the robust statistical regularities found in the French data, in particular the logarithmic dependence on distance of the spatial variogram of prices. Indeed this is a signature of two-dimensional diffusing fields driven by random noise, captured by our stylized model, Eq.~\eqref{eq: diffusion equation}, which was already used in the past to model spatial regularities in voting patterns \cite{Borghesi2010, borghesi2012election}. Note that a model where prices propagate in a  ballistic way ($r \sim t$) instead of diffusing ($r \sim \sqrt{t}$) would lead to completely different spatial correlations. 
The temporal fluctuations of prices can be accounted for within the same framework, provided the shocks are persistent over a time scale that we find to be around 3 years. The data also suggests, not surprisingly, that the amplitude of the price shocks is spatially heterogeneous, with a large variation span. The order of magnitude of the diffusion constant was estimated through an impulse response analysis. Other dimensional parameters obtained from fitting the spatial and temporal correlations appear to be of reasonable order of magnitude.  

Our study thus quantifies the diffusive nature of housing prices that was anticipated long ago \cite{clapp1994, Pollakowski1997}, albeit on more restricted, local data sets. The possibility of describing the spatio-temporal dynamics of housing prices is clearly interesting from many standpoints, in particular for land use planning and territorial development policies or for real estate investment strategies as a response to ``shocks'', like the opening of a fast train or a metro station. Future work should attempt couple the random diffusion equation for prices to the population field in order to describe social mobility, as a two-field extension of our previous work \cite{Zakine2023}. Extending our analysis to other spatial socio-economic variables would also shed light on the mechanisms underlying diffusion of socio-cultural traits, as suggested in \cite{bouchaud2014emergence}.   

\paragraph{Acknowledgements} We thank Xavier Gabaix, Swann Chelly, Nirbhay Patil and Max Sina Knicker for fruitful comments and discussions. We also thank Thomas Piketty for useful explanations about how the data published in \cite{Piketty2023} was created. This research was conducted within the Econophysics $\&$ Complex Systems Research Chair, under the aegis of the Fondation du Risque, the Fondation de l’École polytechnique, the École polytechnique and Capital Fund Management.

\bibliographystyle{apsrev4-2}

\begin{thebibliography}{0}%
\makeatletter
\providecommand \@ifxundefined [1]{%
 \@ifx{#1\undefined}
}%
\providecommand \@ifnum [1]{%
 \ifnum #1\expandafter \@firstoftwo
 \else \expandafter \@secondoftwo
 \fi
}%
\providecommand \@ifx [1]{%
 \ifx #1\expandafter \@firstoftwo
 \else \expandafter \@secondoftwo
 \fi
}%
\providecommand \natexlab [1]{#1}%
\providecommand \enquote  [1]{``#1''}%
\providecommand \bibnamefont  [1]{#1}%
\providecommand \bibfnamefont [1]{#1}%
\providecommand \citenamefont [1]{#1}%
\providecommand \href@noop [0]{\@secondoftwo}%
\providecommand \href [0]{\begingroup \@sanitize@url \@href}%
\providecommand \@href[1]{\@@startlink{#1}\@@href}%
\providecommand \@@href[1]{\endgroup#1\@@endlink}%
\providecommand \@sanitize@url [0]{\catcode `\\12\catcode `\$12\catcode `\&12\catcode `\#12\catcode `\^12\catcode `\_12\catcode `\%12\relax}%
\providecommand \@@startlink[1]{}%
\providecommand \@@endlink[0]{}%
\providecommand \url  [0]{\begingroup\@sanitize@url \@url }%
\providecommand \@url [1]{\endgroup\@href {#1}{\urlprefix }}%
\providecommand \urlprefix  [0]{URL }%
\providecommand \Eprint [0]{\href }%
\providecommand \doibase [0]{https://doi.org/}%
\providecommand \selectlanguage [0]{\@gobble}%
\providecommand \bibinfo  [0]{\@secondoftwo}%
\providecommand \bibfield  [0]{\@secondoftwo}%
\providecommand \translation [1]{[#1]}%
\providecommand \BibitemOpen [0]{}%
\providecommand \bibitemStop [0]{}%
\providecommand \bibitemNoStop [0]{.\EOS\space}%
\providecommand \EOS [0]{\spacefactor3000\relax}%
\providecommand \BibitemShut  [1]{\csname bibitem#1\endcsname}%
\let\auto@bib@innerbib\@empty
\end{thebibliography}%


\begin{thebibliography}{99}



\bibitem{clapp1994}
J. M. Clapp and D. Tirtiroglu, "Positive feedback trading and diffusion of asset price changes: Evidence from housing transactions," J. Econ. Behav. Organ. \textbf{24}(3), 337-355 (1994).

\bibitem{Pollakowski1997}
H. O. Pollakowski and T. S. Ray, "Housing Price Diffusion Patterns at Different Aggregation Levels," J. Housing Res. \textbf{8}(1), 107-124 (1997).

\bibitem{barabasi1995fractal}
A.-L. Barabási and H. E. Stanley, \textit{Fractal concepts in surface growth} (Cambridge University Press, 1995).

\bibitem{frisch1995turbulence}
U. Frisch, \textit{Turbulence: the legacy of A. N. Kolmogorov} (Cambridge University Press, 1995).

\bibitem{Borghesi2010}
C. Borghesi and J.-P. Bouchaud, "Spatial correlations in vote statistics: a diffusive field model for decision-making," Eur. Phys. J. B \textbf{75}, 395-404 (2010).

\bibitem{borghesi2012election}
C. Borghesi, J.-C. Raynal, and J.-P. Bouchaud, "Election turnout statistics in many countries: a diffusive field model for decision-making," PLoS ONE \textbf{7}(5), e36289 (2012).

\bibitem{fernandez2014voter}
J. Fernández-Gracia et al., "Is the voter model a model for voters?" Phys. Rev. Lett. \textbf{112}(15), 158701 (2014).

\bibitem{geanakoplos2012getting}
J. Geanakoplos et al., "Getting at systemic risk via an agent-based model of the housing market," Am. Econ. Rev. \textbf{102}(3), 53-58 (2012).

\bibitem{Rosen1974}
S. Rosen, "Hedonic Prices and Implicit Markets: Product Differentiation in Pure Competition," J. Polit. Econ. \textbf{82}(1), 34-55 (1974).

\bibitem{Piketty2023}
J. Cagé and T. Piketty, \textit{Une histoire du conflit politique}, Le Seuil (2023).

\bibitem{Besley2012}
T. Besley and H. Mueller, "Estimating the Peace Dividend: The Impact of Violence on House Prices in Northern Ireland," Am. Econ. Rev. \textbf{102}(2), 810-833 (2012).

\bibitem{Figlio2004}
D. N. Figlio and M. E. Lucas, "What's in a Grade? School Report Cards and the Housing Market," Am. Econ. Rev. \textbf{94}(3), 591-604 (2004).



\bibitem{Feitosa2008}
F. Feitosa and W. Zesk, "Spatial Patterns of Residential Segregation: A Generative Model," Working Paper (2008).

\bibitem{Pangallo2019}
M. Pangallo, J.-P. Nadal, and A. Vignes, "Residential income segregation: A behavioral model of the housing market," J. Econ. Behav. Organ. \textbf{159}, 15-35 (2019).

\bibitem{Baptista2016}
R. Baptista, J. D. Farmer, and A. Uluc, "Macroprudential policy in an agent-based model of the UK housing market," Bank of England Staff Working Paper No. 619 (2016).

\bibitem{gauvin_modeling_2013}
L. Gauvin, A. Vignes, and J.-P. Nadal, "Modeling urban housing market dynamics: Can the socio-spatial segregation preserve some social diversity?" J. Econ. Dyn. Control \textbf{37}(7), 1300-1321 (2013).



\bibitem{samuelson2016proof}
P. A. Samuelson, "Proof that properly anticipated prices fluctuate randomly," in \textit{The World Scientific Handbook of Futures Markets}, World Scientific (2016).


\bibitem{Thibodeau1998}
S. Basu and T. Thibodeau, "Analysis of Spatial Autocorrelation in House Prices," J. Real Estate Finance Econ. \textbf{17}(1), 61-85 (1998).

\bibitem{kelly2019}
T.G. Conley, M. Kelly, "The Standard Errors of Persistence", Journal of International Economics, Volume 153, (2025).

\bibitem{schweitzer2000modelling}
F. Schweitzer and J. A. Ho{\l}yst, "Modelling collective opinion formation by means of active Brownian particles," Eur. Phys. J. B \textbf{15}, 723-732 (2000).

\bibitem{schweitzer2004coordination}
F. Schweitzer, "Coordination of decisions in a spatial model of Brownian agents," in \textit{The Complex Dynamics of Economic Interaction}, Springer (2004).

\bibitem{bouchaud2014emergence}
J.-P. Bouchaud, C. Borghesi, and P. Jensen, "On the emergence of an ‘intention field’ for socially cohesive agents," J. Stat. Mech. \textbf{2014}(3), P03010 (2014).

\bibitem{short2010} M. B. Short, P. J. Brantingham, A. L. Bertozzi, and G. E. Tita, ``Dissipation and displacement of hotspots in reaction-diffusion models of crime," Proceedings of the National Academy of Sciences, 107(9), 3961-3965 (2010).

\bibitem{edwards1982surface}
S. F. Edwards and D. R. Wilkinson, "The surface statistics of a granular aggregate," Proc. R. Soc. Lond. A \textbf{381}(1780), 17-31 (1982).

\bibitem{Anderson1972}
P. W. Anderson, "More Is Different," Science \textbf{177}(4047), 393-396 (1972).

\bibitem{bach2015top}
S. Bach, A. Thiemann, and A. Zucco, "The top tail of the wealth distribution in Germany, France, Spain, and Greece," DIW Berlin Discussion Paper (2015).

\bibitem{Zakine2023}
R. Zakine, J. Garnier-Brun, A.-C. Becharat, and M. Benzaquen, "Socioeconomic agents as active matter in nonequilibrium Sakoda-Schelling models," Phys. Rev. E \textbf{109}(4), 044310 (2024).


\end{thebibliography}

\onecolumngrid

\section*{\bf{Supplemental Material}}

\setcounter{secnumdepth}{1}
\section{SM-1: Analytical derivation of the diffusive term}
\label{sec:SM1}

We assume that the diffusive term in the price field evolves through a mechanism of supply and demand such that the time evolution of the field $\psi$ depends on the difference of the field between two locations $\psi(R_\alpha) -\psi(R_\beta)$ where $R_\alpha$ and $R_\beta$ refer to the considered locations.
We then propose the following generic equation to describe the propagation of the field with respect to its surrounding influences:
\begin{equation}
    \partial_t \psi(R_\alpha, t)= \sum_\beta \Gamma_{\alpha,\beta} \psi(R_\beta) - \sum_\beta \Gamma_{\beta,\alpha} \psi(R_\alpha),
\end{equation} where $\Gamma$ is a symmetric influence matrix such that: \begin{equation}
    \Gamma_{\alpha,\beta} = \Gamma(R_\alpha \vert R_\beta)=t(R_\alpha-R_\beta \vert R_\beta ).
\end{equation}
Hence, in the continuous limit and in one dimension for simplicity, it comes: \begin{equation}
    \partial_t \psi(x,t)= \int t(x-x' \vert x')\psi(x',t)dx' - \int  t(x'-x \vert x)\psi(x,t)dx',
\end{equation} which we can re write as: \begin{equation}
    \partial_t \psi(x,t)= \int t(y \vert x-y)\psi(x-y,t)dy - \int t(y \vert x)\psi(x,t)dy,
    \label{eq : Master equation}
\end{equation}
changing variables to $y=x-x'$.
The Kramers-Moyal expansion of \eqref{eq : Master equation} up to the order 2 in $y$ then gives: \begin{equation}
    \partial_t \psi(x,t)= - \partial_x \left[R_1(x) \psi(x)\right] + \frac{1}{2} \partial^2_x \left[R_2(x) \psi(x)\right],
\end{equation}
where:
    \begin{align}
        R_1(x) = \int  y t(y,x) dy;\\
        R_2(x) = \int y^2 t(y,x) dy.
    \end{align}
Moreover, the influence matrix is symmetric, hence the drift term $R_1(x)$ is set to zero and we retrieve the one dimensional diffusion equation: \begin{equation}
        \partial_t \psi(x,t)=  \partial^2_x \left[D(x) \psi(x)\right]
\end{equation} with $D(x) = \frac{1}{2} \int  y^2 t(y,x) dy$.
Note that we retrieve here a non-uniform diffusion coefficient, but we assume in the rest of the study that we can take $D(x)=D$.

\section{SM-2: Theoretical predictions for the variograms}
\label{sec:SM2}

\subsection{SM-2.1: Computation of the generic space-time variogram}

Let us consider the following stochastic partial differential equation: \begin{equation}
    \frac{\partial \psi(\bf{r},t)}{\partial t} = D \Delta \psi(\bf{r},t) - \varkappa \psi(\bf{r},t) + \eta (\bf{r},t) + \xi(\bf{r}),
\end{equation} where $\Delta$ is the Laplacian operator, $D$  a diffusion coefficient, $\varkappa$ a mean-reversion coefficient, $\eta (\bf{r},t)$  a Langevin noise with zero mean and short range time and space correlations, and $\xi(\bf{r})$  a static random field with zero mean and short range correlations. The correlators of these terms are assumed to be of the following type:
\begin{align} \nonumber
    \left \langle \eta(\bf{r},t) \eta(\bf{r'},t') \right \rangle &= \frac{A^2}{T a^2} e^{-|t-t'|/T} g_a(|\bf{r} - \bf{r'}|); \\
    \left \langle \xi(\bf{r}) \xi(\bf{r'}) \right \rangle &= \frac{\Sigma^2}{a^2} g_a(|\bf{r} - \bf{r'}|),
\end{align}
where $g_a(r)$ is a  bell-shaped function that decays over length scale $a$, such that $2 \pi \int_{r \geq 0} g_a(r) r {\rm d}r = a^2$.
For the rest of the calculations, we consider the regime where $|\mathbf{r} - \mathbf{r'}| = \ell \gg a$
 which leads to $\frac{1}{a^2} g_a(|\bf{r}-\bf{r'}|) \approx  \delta(|\bf{r}-\bf{r'}|)$.
Moreover, the space time correlation function can be written as: \begin{equation}
    \mathbb{C}(|\mathbf{r}-\mathbf{r'}|,|t- t'|) = \langle\psi(\mathbf{r},t)\psi(\mathbf{r'},t')\rangle = \int \int e^{-i \mathbf{k}\mathbf{r}-i \mathbf{k'}\mathbf{r'}} \langle\psi_\mathbf{k}(t)\psi_\mathbf{k'}(t')\rangle \frac{d\mathbf{k}}{(2\pi)^2}\frac{d\mathbf{k'}}{(2\pi)^2},
\end{equation} where $\psi_\mathbf{k}$ is the solution of the following equation in Fourier space: \begin{equation}
    \frac{\partial \psi_\mathbf{k}(t)}{\partial t} = -D \mathbf{k}^2 \psi_\mathbf{k}(t)-\varkappa \psi_\mathbf{k}(t) + \eta_\mathbf{k} + \xi_\mathbf{k}.
\end{equation}
Hence: \begin{equation}
    \psi_\mathbf{k}(t) = \psi_\mathbf{k}(0) e^{-(D\mathbf{k}^2+\varkappa) t} +  \int_{0}^{t} e^{-(D\mathbf{k}^2+\varkappa) (t-\tau)}(\eta_\mathbf{k}(\tau)+\xi_\mathbf{k} )d\tau.
\end{equation}
Because of the two fields $\eta$ and $\xi$ - assumed to be independent - we will separate the calculation for the correlation function into two contributions.
 In the long time limit, the first contribution in Fourier space, coming from field $\eta$, is: \begin{equation}
      \int_{0}^{t} \int_{0}^{t'} dt_1 dt_2 e^{-(D\mathbf{k}^2+\varkappa)(t-t_1) -(D\mathbf{k'}^2+\varkappa)(t'-t_2)}\langle\eta_\mathbf{k}(t_1)\eta_\mathbf{k'}(t_2)\rangle,
\end{equation}
leading to:
\begin{equation}
    \frac{A^2(2\pi)^2}{T} \int_{0}^{t} \int_{0}^{t'} dt_1 dt_2 e^{-(D\mathbf{k}^2+\varkappa)(t-t_1) -(D\mathbf{k'}^2+\varkappa)(t'-t_2)}e^{-\frac{\abs{t_1-t_2}}{T}}\delta(\mathbf{k}+\mathbf{k'}).
\end{equation}
We find, in the long time limit, that the integral yields in Fourier space:
\begin{equation}
\begin{split}
    \frac{A^2(2\pi)^2}{2T}\left[ \frac{e^{-(D\mathbf{k}^2+\varkappa)\abs{t'-t}}}{2(D\mathbf{k}^2+\varkappa)(D\mathbf{k}^2+\varkappa+\frac{1}{T})}  
    +\frac{e^{-\frac{\abs{t-t'}}{T}}}{(D\mathbf{k}^2+\varkappa)^2 -\frac{1}{T^2}} - \frac{e^{-(D\mathbf{k}^2+\varkappa)\abs{t'-t}}}{2(D\mathbf{k}^2+\varkappa)(D\mathbf{k}^2+\varkappa-\frac{1}{T})}\right].
    \end{split}
\end{equation}
This can be condensed as:
\begin{equation}
  \frac{A^2(2\pi)^2}{2T((D\mathbf{k}^2+\varkappa)^2 -\frac{1}{T^2})}\left[e^{-\frac{\abs{t-t'}}{T}}-\frac{e^{-(D\mathbf{k}^2+\varkappa)\abs{t'-t}}}{T(D\mathbf{k}^2+\varkappa)}\right].
    \label{eq : temporal correl exponentials}
\end{equation}
Similarly, we can compute the contribution for the correlation function coming from field $\xi(\bf{r})$:
\begin{equation}
     (2 \pi)^2 \Sigma^2   \int_{0}^{t} \int_{0}^{t'} dt_1 dt_2 e^{-(D\mathbf{k}^2+\varkappa)(t-t_1) -(D\mathbf{k'}^2+\varkappa)(t'-t_2)}\delta(\mathbf{k}+\mathbf{k'}) .
\end{equation}
This yields, in the long time limit:
\begin{equation}
  \frac{(2 \pi)^2 \Sigma^2}{ (D\mathbf{k}^2+\varkappa)^2}. 
\end{equation}
In the next sections, we will show how, starting from what has just been shown, we compute both the spatial and the temporal variograms, defined as $\mathbb{V}(\ell,0) := \langle (\psi({\bf r},t) - \psi({\bf r}',t))^2 \rangle$ and $\mathbb{V}(0,\tau) := \langle (\psi({\bf r},t) - \psi({\bf r},t + \tau))^2 \rangle$.

\subsection{SM-2.2: Computation of the spatial variogram}

We come back to the first contribution (coming from field $\eta$) in Fourier space for the space time correlation function: 
\begin{equation}
    \frac{A^2(2\pi)^2}{2T((D\mathbf{k}^2+\varkappa)^2 -\frac{1}{T^2})}\left[e^{-\frac{\abs{t-t'}}{T}}-\frac{e^{-(D\mathbf{k}^2+\varkappa)\abs{t'-t}}}{T(D\mathbf{k}^2+\varkappa)}\right].
    \end{equation}
We now focus on the static behavior of this term, hence imposing $t=t'$.
This yields: 
\begin{equation}
    \frac{A^2(2\pi)^2}{2T((D\mathbf{k}^2+\varkappa)^2 -\frac{1}{T^2})}\left[1-\frac{1}{T(D\mathbf{k}^2+\varkappa)}\right].
    \end{equation}
Using notations $\vert \mathbf{k} \vert = k $, $\mathbf{k}.(\mathbf{r}-\mathbf{r'}) = k \ell \cos(\theta)$ and notation $\mathbb{C}_{\eta}$ to describe the contribution from $\eta$ to the correlation function, it comes in polar coordinates:
\begin{equation}
    \mathbb{C}_{\eta}(\ell,0) = \frac{A^2}{2T (2\pi)^2} \int dk \int d\theta e^{-i k \ell \cos(\theta)}\frac{k}{((Dk^2+\varkappa)^2 -\frac{1}{T^2})}\left[1-\frac{1}{T(Dk^2+\varkappa)}\right].
\end{equation} 
The integral is defined for $ 1/\ell^* \ll k \ll 1/a$, which ensures that
$Dk^2 \gg \frac{D}{\ell^{*2}}=\varkappa$. We can hence neglect the mean-reversion term in the computation. Moreover, we can neglect $D^2 k^4$ in favor of $\frac{1}{T^2}$ if $Dk^2 < \frac{1}{T}$, hence if $\ell > \sqrt{DT}$. This is typically the regime that we consider for this study, since we estimate (see in the main text) $\sqrt{DT} \approx 13$ km, so we assume here that this term is negligible.
Finally, we can identify the Bessel function \begin{equation}
    \frac{1}{ 2\pi}\int_{0}^{2 \pi} d\theta e^{i k \ell \cos(\theta)} = J_0(k \ell)=J_0(-k\ell),
\end{equation} 
so:
\begin{equation}
    \mathbb{C}_{\eta}(\ell,0)  \approx \frac{A^2}{4 \pi} \int_{1/\ell^*}^{1/a} dk \frac{J_0(k \ell)}{Dk}.
\end{equation}
The Bessel function can be expanded for $k \ell \longrightarrow 0$, and yields $J_0(k \ell) \approx 1-\ell^2k^2 / 4 + o(k^4 \ell^4)$. Moreover, the Bessel function decays to zero when $k \ell \gg 1$, concentrating the integral towards its lower bound. This gives, up to constant contributions:
\begin{equation}
    \mathbb{C}_{\eta}(\ell,0)  \approx  -\frac{A^2}{4 \pi D}\log \frac{\ell}{\ell^*} + K(\ell)
\end{equation} with correction term $K(\ell)$.
Similarly, we can compute the contribution from field $\xi$:
\begin{equation}
    \mathbb{C}_{\xi}(\ell,0)  = \frac{\Sigma^2}{2 \pi D^2} \int_{1/\ell^*}^{1/a} dk \frac{J_0(k \ell)}{k^3} = \frac{\Sigma^2}{2 \pi D^2}\ell^2 \int_{\ell/\ell^*}^{\ell/a} du \frac{J_0(u)}{u^3}.
\end{equation}
In order to have a non-constant contribution here, we must go to the second order in the expansion of the Bessel function towards the lower bound of the integral. This yields:
\begin{equation}
    \mathbb{C}_{\xi}(\ell,0)  \approx \frac{\Sigma^2}{2 \pi D^2}\ell^2 \int_{\ell/\ell^*}^{\ell/a} du \frac{1-\frac{u^2}{4}}{u^3},
\end{equation}
which finally yields, up to constant terms:
\begin{equation}
    \mathbb{C}_{\xi}(\ell,0)  \approx  \frac{\Sigma^2}{8 \pi D^2} \ell^2 \log \frac{\ell}{\ell^*}+ K'(\ell)
\end{equation} with correction $K'(\ell)$.
Furthermore, the variogram is defined as $\mathbb{V}(\ell,0) = 2 \langle \psi(\mathbf{r},0)^2 \rangle - 2\mathbb{C}(\ell,0)$.
Hence, summing both contributions yields:
\begin{equation}
   \mathbb{V}(\ell,0) \approx \frac{A^2}{2 \pi D}\log \frac{\ell}{\ell^*} -\frac{\Sigma^2}{4 \pi D^2}\ell^2\log \frac{\ell}{\ell^*} + C,
\end{equation}
where $C$ is a constant.
This result is of course only valid in the range where $a \ll \ell \ll \ell^*$.

\subsection{SM-2.3: Computation of the temporal variogram}

As we are now interested in the temporal variation of the same point in space, we will neglect the random static field $\xi(\bf{r})$ in the computation which will only yield constant terms. Moreover, we will again neglect the contribution $\varkappa$ in the calculations as the integration back to real space will impose $Dk^2 \gg \varkappa$, as seen in the previous section.
Our starting point is therefore the following:
\begin{equation}
    \frac{A^2(2\pi)^2}{2T(D^2\mathbf{k}^4 -\frac{1}{T^2})}\left[e^{-\frac{\abs{t-t'}}{T}}-\frac{e^{-D\mathbf{k}^2\abs{t'-t}}}{TD\mathbf{k}^2}\right].
    \end{equation}
The computation will yield different results depending on the relative values of $\tau=\abs{t-t'}$ and $T$.

When $\tau=\abs{t-t'} \gg T$, we can set $e^{-\frac{\abs{t-t'}}{T}}$ to zero.
Coming back in real space yields:  \begin{equation}
    \mathbb{C}(0,\vert t-t' \vert)  =- \frac{A^2}{T^2(2 \pi)^2} \int d\mathbf{k} \frac{ e^{-D \mathbf{k}^2 \abs{t-t'}}}{2 D\mathbf{k}^2(D^2\mathbf{k}^4 -\frac{1}{T^2})},   
\end{equation} which gives in polar coordinates:
\begin{equation}
    \mathbb{C}(0,\tau)  = -\frac{A^2}{T^2(2 \pi)^2} \int dk \int d\theta \frac{k e^{-D k^2 \tau}}{2Dk^2 (D^2k^4 -\frac{1}{T^2})}.
\end{equation}
It comes:
\begin{equation}
    \mathbb{C}(0,\tau) = -\frac{ A^2}{8 \pi D T^2}\int_{\frac{D \tau}{\ell^{*2}}}^{\frac{D \tau}{a^2}} du  \frac{e^{-u}}{u (\frac{u^2}{\tau^2} - \frac{1}{T^2})}.
    \label{eq : correlator eta}
\end{equation}
Moreover, $\frac{u}{\tau} < \frac{1}{T}$ if $S= \frac{a^2}{D} > T$,
which allows us to neglect this term, leading to:
\begin{equation}
    \mathbb{C}(0,\tau) \approx \frac{ A^2}{8 \pi D }\int_{\frac{D \tau}{\ell^{*2}}}^{\frac{D \tau}{a^2}} du  \frac{e^{-u}}{u}.
\end{equation}
Hence, in the regime where  $T < S \ll \tau \ll \varkappa^{-1}= \frac{\ell^{*2}}{D}$:
\begin{equation}
    \mathbb{C}(0,\tau) \approx - \frac{ A^2}{8 \pi D} \log \tau,
\end{equation}
up to constant terms.
This finally yields:
\begin{equation}
    \mathbb{V}(0,\tau)  \approx  \frac{ A^2}{4 \pi D} \log \tau .
\end{equation}
When $S \ll T \ll \tau \ll \varkappa^{-1}$, logarithmic contributions can once again be obtained by performing a partial fraction decomposition in \eqref{eq : correlator eta} prior to integration.
For completeness, in the regime where $\tau \gg \varkappa^{-1}, S, T$, the computation yields a constant value. 

When $\tau=\abs{t-t'} \ll T$, we come back to:  
\begin{equation}
    \frac{A^2(2\pi)^2}{2T(D^2\mathbf{k}^4 -\frac{1}{T^2})}\left[e^{-\frac{\abs{t-t'}}{T}}-\frac{e^{-D\mathbf{k}^2\abs{t'-t}}}{TD\mathbf{k}^2}\right].
    \end{equation}
If $ \tau \ll S$, we can expand up to the order two in the exponentials for $D \mathbf{k}^2\tau \longrightarrow 0$, in addition to the expansion for $\frac{\tau}{T} \longrightarrow 0$, leading to:
\begin{equation}
    \frac{A^2(2\pi)^2}{2T(D^2\mathbf{k}^4 -\frac{1}{T^2})}\left[\frac{TD\mathbf{k}^2-1}{TD\mathbf{k}^2} + \frac{1}{2}(1-TD\mathbf{k}^2) \frac{\tau^2}{T^2}\right].
    \label{eq : correlator fourier with constant term}
\end{equation}
Hence, the temporal contribution in the correlation function, coming back to real space, is:
\begin{equation}
   \mathbb{C}(0,\tau) = \frac{1}{2 \pi} \int_{\frac{1}{\ell^*}}^{\frac{1}{a}} dk \frac{A^2 k}{2T\left(D^2k^4 -\frac{1}{T^2}\right)} \frac{1}{2}(1-TD k^2) \frac{\tau^2}{T^2}.
\end{equation}
This yields, after integration and up to constant terms:
\begin{equation}
    \mathbb{C}(0,\tau) \approx \frac{ A^2}{32\pi D} \log\left(\frac{\frac{TD}{\ell^{*2}}+1}{TD/a^2 +1}\right)\frac{\tau^2}{T^2},
\end{equation}
which we can re write as: 
\begin{equation}
    \mathbb{C}(0,\tau)\approx -\frac{A^2}{32 \pi D} \log\left(\frac{\frac{T}{S} +1}{\varkappa T+1}\right)\frac{\tau^2}{T^2}.
\end{equation}
This finally yields:
\begin{equation} 
    \mathbb{V}(0,\tau)\approx \frac{ A^2}{16 \pi D} \log\left(\frac{\frac{T}{S} +1}{\varkappa T+1}\right)\frac{\tau^2}{T^2}.
\label{eq : variogram temporel tau petit devant S}
\end{equation}
If $\tau \geq S$, we cannot expand in the second exponential term of \eqref{eq : temporal correl exponentials}. This leads us to study separately both terms. The first one will give, after expanding up to the second order in $\frac{\tau}{T}$: 
\begin{equation}
 \frac{1}{2 \pi}  \int k dk \frac{A^2}{2T(D^2 k^4 -\frac{1}{T^2})}\left(1-\frac{\tau}{T} + \frac{\tau^2}{2T^2}\right),
\end{equation}
which yields: 
\begin{equation}
    \frac{ A^2}{32 \pi D} \log\left(\frac{\abs{\frac{T}{S} -1}(\varkappa T +1)}{(\frac{T}{S} +1)\abs{\varkappa T-1}}\right)\left(1- \frac{\tau}{T}+\frac{\tau^2}{2T^2}\right).
\end{equation}
The second term:
\begin{equation}
    -\frac{1}{2 \pi} \frac{A^2}{2T(D^2\mathbf{k}^4 -\frac{1}{T^2})}\frac{e^{-D\mathbf{k}^2\abs{t'-t}}}{TD\mathbf{k}^2}
\end{equation}
will give:
\begin{equation}
   - \frac{ A^2}{8 \pi T^2D} \int_{\frac{1}{\ell^*}}^{\frac{1}{a}} dk \frac{e^{-Dk^2 \tau}}{k (Dk^2-\frac{1}{T})(Dk^2+\frac{1}{T})}.
\end{equation}
Changing variables to $u=Dk^2\tau$ yields, after a few integration steps:
\begin{equation}
    -\frac{ A^2 }{32 \pi D} \left[e^{\tau/T} \log\left(\frac{\abs{\frac{T}{S} -1}}{\abs{\varkappa T-1}}\right) + e^{-\tau/T} \log \left(\frac{\frac{T}{S}+1}{\varkappa T+1}\right) - 2 \log\left(\varkappa S \right)\right],
\end{equation}
which gives, after expanding the two exponentials $e^{\tau/T}$ and $e^{-\tau/T}$ up to the order two in $\frac{\tau}{T}$:
\begin{equation}
     -\frac{A^2}{32 \pi D} \log\left( \frac{\abs{\frac{T}{S} -1}(\varkappa T +1)}{(\frac{T}{S} +1)\abs{\varkappa T-1}}\right)  \frac{\tau}{T} -\frac{ A^2}{ 64 \pi D} \log\left(\frac{\abs{\frac{T^2}{S^2}-1}}{\abs{\varkappa^2 T^2-1}}\right) \frac{\tau^2}{T^2}.
\end{equation}
This finally yields, after adding the first and second term contribution from \eqref{eq : temporal correl exponentials}:
\begin{equation}
    \mathbb{V}(0,\tau) \approx   \frac{ A^2}{8 \pi D} \log\left(\frac{\abs{\frac{T}{S} -1}(\varkappa T +1)}{(\frac{T}{S} +1)\abs{\varkappa T-1}}\right)  \frac{\tau}{T} + \frac{ A^2}{16 \pi D} \log\left(\frac{\frac{T}{S} +1}{\varkappa T+1}\right)\frac{\tau^2}{T^2}.
\end{equation}
We hence lose the quadratic behavior for the variogram when $S \leq \tau \ll T$ and the dominant behavior becomes linear.

\newpage
\section{SM-3: Additional Plots}
\label{sec:SM3}

\begin{figure}[h]
    \centering
    \includegraphics[width=0.5 \linewidth]{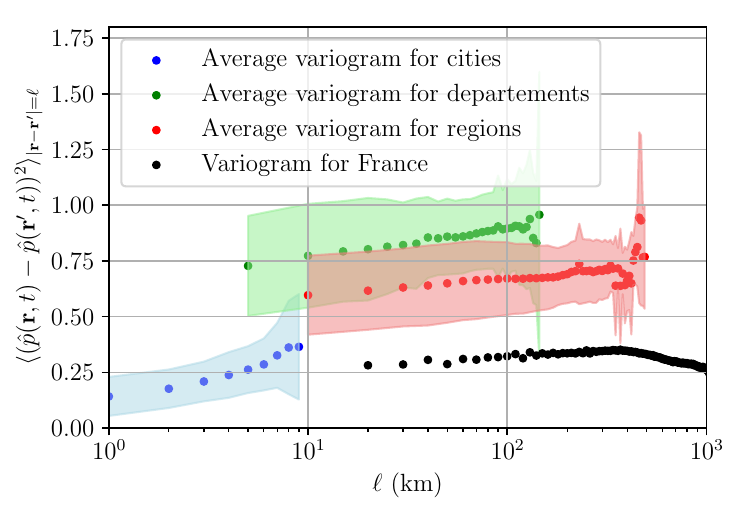}
    \caption{Spatial variogram for the log-price field per squared meter $\hat{p}({\bf r}):=\log(\hat{P})$, where the notation $\hat{P}$ indicates the prices per squared meter, averaged over the period 2018-2022 for France as a whole, its \textit{régions}, \textit{départements} and cities, with their respective cross-sectional variability highlighted in shaded colors and the averages for each scale as filled circles. The different off-sets in the $y$ direction corresponds to the measurement noise contribution to the empirical field $\hat{p}({\bf r})$. The observed empirical behavior is once again logarithmic, with slopes ranging between 0.03 and 0.06, with an average value approximately half of that found in \cite{Borghesi2010}.}
    \label{fig:variograms m2}
\end{figure}

\begin{figure}[h]
    \centering
        \includegraphics[width=0.5 \linewidth]{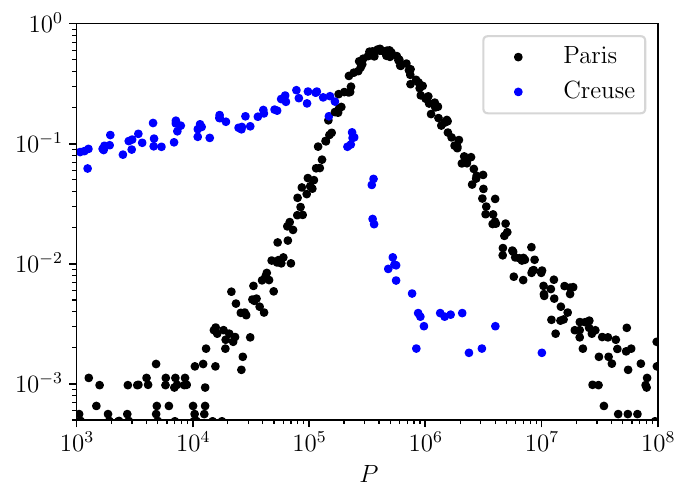} 
    \caption{Distribution of all transaction log-prices ${p}:=\log {P}$, averaged for the 5 years of DVF data, both for the \textit{département} of la Creuse and for Paris. These locations were chosen as typical examples of both the countryside and cities, showing clearly two different shapes. This explains the double-hump nature of the global log-price distribution for the whole of France, discussed in the main text.}
    \label{fig:distribution log price exemples}
\end{figure}

\begin{figure}[h]
    \centering
    \includegraphics[width=0.5 \linewidth]{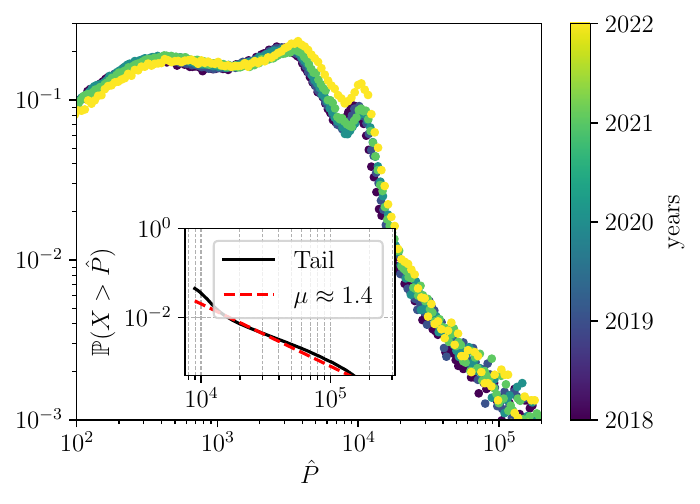}
    \caption{Distribution of all transaction log-prices per squared meter $\hat{p}$, for the 5 years of DVF data. The right tail corresponds to a power-law tail for prices per squared meter as $\hat{P}^{-1-\mu}$ with $\mu \approx 1.4$, close to 1.5, as found for the log-prices above.}
    \label{fig:dsitribution m2}
\end{figure}

\begin{figure}[h]
    \centering
    \includegraphics[width=0.5 \linewidth]{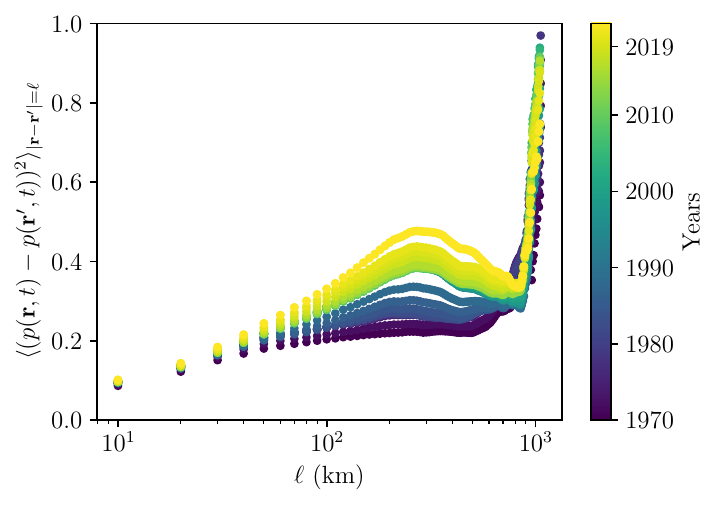}
    \caption{Spatial variograms for the log price field $p(\mathbf{r})$ for every year between 1970 to 2022, using the data from \cite{Piketty2023}. We see that the slope of these variograms is only weakly time-dependent, and that the logarithmic behavior is robust in time. The variogram saturates for $\ell \approx 70$ km in 1970 and for $\ell \approx 300$ km in 2022.}
    \label{fig:vario piketty}
\end{figure}

\begin{figure}[h]
    \centering
    \includegraphics[width=0.5 \linewidth]{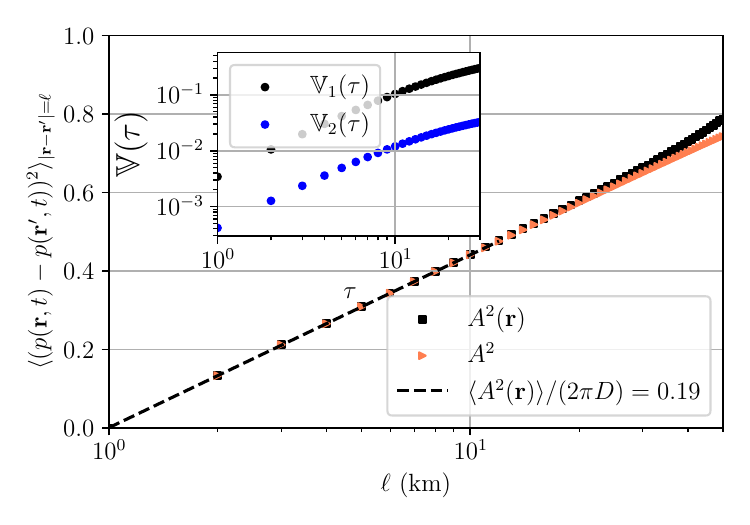}
    \caption{Theoretical predictions for the spatial variogram, computed when the noise amplitude is uniform, equal to $A^2$, and when the noise amplitude $A^2(\mathbf{r})$ is strongly heterogeneous, with $\langle A^2(\mathbf{r}) \rangle = A^2 = 2 \pi D \times 0.19$. 
    We obtain a similar logarithmic behavior in both cases. The inset shows a comparison between $\mathbb{V}_1(\tau)$ and $\mathbb{V}_2(\tau)$ computed for data simulated on a lattice with the same strongly heterogeneous noise amplitude $A^2(\mathbf{r})$. We hence qualitatively retrieve the observed empirical temporal behavior.}
    \label{fig:msd simu}
\end{figure}

\FloatBarrier
\clearpage
\section{SM-4: Estimating the diffusion constant $D$}
\label{sec:SM4}

To estimate the order of magnitude of the diffusion constant \( D \) in France, we examine the propagation of price ``shocks'' induced by the opening of a TGV ({\it Train à Grande Vitesse}) station in several cities (Lyon, Bordeaux, and Tours) and their surrounding areas. For each area, we compute

\[
\sigma^2(t) = \frac{\sum  R(\textbf{r}, t) (\textbf{r}-\bar{\textbf{r}})^2 }{\sum R(\textbf{r}, t)},
\]

where \( R(\textbf{r}, t) = p(\textbf{r}, t) - p(\textbf{r}, t_0) \), with the summation taken over all \textit{communes} within the considered areas. Our findings indicate that, for these three regions (indexed by \( i \)), the relation 

\[
\sigma_i^2(t) \approx D_i t + C_i
\]

holds, allowing us to estimate an order of magnitude for the diffusion constant from the slopes of the corresponding curves, as illustrated in Fig.~4 of the main text.

\end{document}